%% file: main.tex
\documentclass[%
amsmath,aps,showkeys,
amssymb,twocolumn,pra,
superscriptaddress
]{revtex4-2}

\usepackage{hyperref}
\hypersetup{
    colorlinks,
    linkcolor={black},
    citecolor={blue!80!black},
    urlcolor={blue!80!black}
}

\usepackage{graphicx}
\usepackage{bm}
\usepackage{bbold}
\usepackage{dcolumn}
\newcolumntype{d}[1]{D{.}{.}{#1}}

\usepackage[dvipsnames]{xcolor}

\newcommand{\abs}[1]{\left\lvert{#1}\right\rvert}

\newcommand{\com}[2]{\left[#1,#2\right]}
\newcommand{\proj}[1]{|#1\rangle\langle #1|} 
\newcommand{\rmi}{\mathrm{i}}

\usepackage{url}
\usepackage{tikz}
\usepackage{pgfplots}
\pgfplotsset{compat=1.14}
\usepackage{booktabs}
\usepackage{pgfplotstable}
\usepackage{float}
\usepackage{subcaption}
 
\usepackage{multirow}
\usepackage{bbold}
\usepackage{listings}

\usepackage{enumerate}
\usepackage{latexsym}
\usepackage{amsfonts, amsmath, bbm, amssymb, amsthm}
\usepackage{epic}
\usepackage{braket}
\usepackage{bbm}
\newcommand{\ketbra}[2]{\ket{#1}\bra{#2}}
\usepackage{colortbl}
\usepackage{xcolor}
\definecolor{lightgray}{gray}{0.9}

\begin{document}

\title{Single Qudit Control in $^{87}$Sr via Optical Nuclear Electric Resonance}

\author{Johannes K. Krondorfer}%
\email{johannes.krondorfer@gmail.com}
\affiliation{ 
Institute of Experimental Physics, Graz University of Technology, Petersgasse 16, 8010 Graz, Austria}

\author{Matthias Diez}
\affiliation{
Institute of Experimental Physics, Graz University of Technology, Petersgasse 16, 8010 Graz, Austria}
\affiliation{Institute of Physics, University of Graz, Universit{\"a}tsplatz 5, 8010 Graz, Austria}

\author{Andreas W. Hauser}%
\email{andreas.w.hauser@gmail.com}
\affiliation{ 
Institute of Experimental Physics, Graz University of Technology, Petersgasse 16, 8010 Graz, Austria}

\date{\today}

\begin{abstract}
Optical nuclear electric resonance (ONER) was recently proposed as a fast and robust single-qubit gate mechanism in $^{87}$Sr. Here, we demonstrate through numerical simulations that ONER can be extended to single-qudit control, addressing multiple one-level hyperfine transitions within the ten-dimensional nuclear-spin manifold. We identify suitable operating regimes and show that ONER enables high-fidelity spin manipulations, with simulated $\pi$-gate fidelities exceeding 99.9\%, while maintaining coherence under realistic parameter fluctuations. These results establish a proof-of-principle for optical qudit control in $^{87}$Sr and delineate practical parameter ranges for future experiments, highlighting ONER as a promising pathway toward high-dimensional quantum information processing.
\end{abstract}


\maketitle

\section{Introduction}
Quantum information processing relies on the precise control of quantum states, traditionally implemented using qubits, which encode information in two-level systems. However, qudits, which utilize multi-level quantum states, can offer significant advantages over qubits, including increased performance per computational unit~\cite{campbell_enhanced_2014, chi_programmable_2022, ringbauer_universal_2022}, reduced circuit complexity~\cite{luo_universal_2014}, a more resource-efficient implementation of quantum algorithms and error-correcting codes~\cite{luo_universal_2014, ringbauer_universal_2022, Nikolaeva2024}, and a higher expressibility of model Hamiltonians in quantum simulation applications~\cite{Zhang2014}. As a result, qudit-based quantum computation has gained increasing interest across various quantum platforms, including superconducting circuits~\cite{Peterer_2015,Svetitsky2014}, trapped ions~\cite{ringbauer_universal_2022,Randall2015}, and neutral atoms~\cite{omanakuttan_qudit_2023, zache_fermion_2023, Deutsch2021qudit, ahmed_coherent_2025}.

Among the various physical implementations, nuclear spin states of alkaline earth atoms in atomic lattices are particularly promising for qubit and qudit control, due to their long-lived qudit states, the theoretically well-understood hyperfine structure~\cite{boyd_nuclear_2007}, their naturally emerging multiple spin states and the ability to precisely control their interactions with external fields~\cite{saffman_quantum_2016, henriet_quantum_2020}. These properties have led to experimental demonstrations of high-fidelity qubit gates~\cite{noguchi_quantum_2011, jenkins_ytterbium_2022, ma_universal_2022, muniz_high_2024}, erasure conversion~\cite{ma_high_2023}, and midcircuit operations~\cite{lis_midcircuit_2023, huie_repetitive_2023, norcia_midcircuit_2023}. 

While many works focus on the lowest lying nuclear spin states for qubit encoding, high spin nuclei offer the possibility for qudit encoding in multiple spin states. The $^{87}$Sr system, featuring a nuclear spin of $I = 9/2$, has been a recent subject of theoretical studies~\cite{omanakuttan_qudit_2023, Deutsch2021qudit}, where quantum control is suggested via magnetic radio fields acting on hyperfine levels in the electronic ground state which are Stark-shifted to enable individual addressability. Selectivity and achievable Rabi frequencies are thus limited by the actual energy splitting and the capability to localize the radio frequency magnetic field.  Recently, the experimental implementation of single qudit control was also realized in $^{87}$Sr via two-photon Raman transitions, addressing the lowest four spin states.~\cite{ahmed_coherent_2025}

While standard Raman methods have successfully been applied for qubit~\cite{barnes_assembly_2022} and qudit control~\cite{ahmed_coherent_2025} in $^{87}$Sr, the reported Rabi frequencies are in the order of 1~kHz, and additional light shifts are necessary to avoid undesired spin mixing. To enhance the achievable Rabi frequencies, we recently proposed optical nuclear electric resonance (ONER) as an alternative approach to nuclear spin manipulation.~\cite{krondorfer_nuclear_2023,krondorfer_optical_2024,krondorfer2025opticalnuclearelectricresonance} ONER leverages the interaction between the nuclear quadrupole moment and an optically induced electric field gradient and hyperfine interaction, enabling high-resolution nuclear spin transitions via amplitude-modulated laser light. Specifically, ONER only requires a single amplitude-modulated laser beam to enable nuclear spin control, simplifying experimental implementation and enhancing localization and individual control compared to magnetic control mechanisms. While previous studies demonstrated ONER for qubit control in $^{87}$Sr~\cite{krondorfer2025opticalnuclearelectricresonance}, achieving Rabi frequencies in the order of tens of kHz, its extension to qudit control remains, up to now, unexplored.

In this article, we demonstrate that ONER can be generalized from the previously studied qubit case to full qudit control across the ten-dimensional nuclear-spin manifold of $^{87}$Sr. This requires addressing multiple closely spaced one-level transitions and mitigating crosstalk, which goes well beyond the two-level scenario. We demonstrate how ONER can be used for precise qudit control across multiple nuclear spin states of $^{87}$Sr, achieving Rabi frequencies in the order of tens of kHz. We compute optimal magnetic field strengths and laser parameters to enable selective, high-fidelity spin transitions. We demonstrate that, in principle, all one-level transitions can be addressed via the ONER method. Furthermore, we assess the stability of ONER-based qudit gates under typical noise sources, showing that single spin flip fidelities exceeding 99.9\% can be achieved at realistic laboratory conditions. Finally, we discuss how ONER is compatible with possible two-qudit gates and non-destructive readout techniques.

\section{ONER as Single-Qudit-Gate in $^{87}\text{Sr}$}
\begin{figure}[!t]
    \centering
    \includegraphics[width=0.48\textwidth]{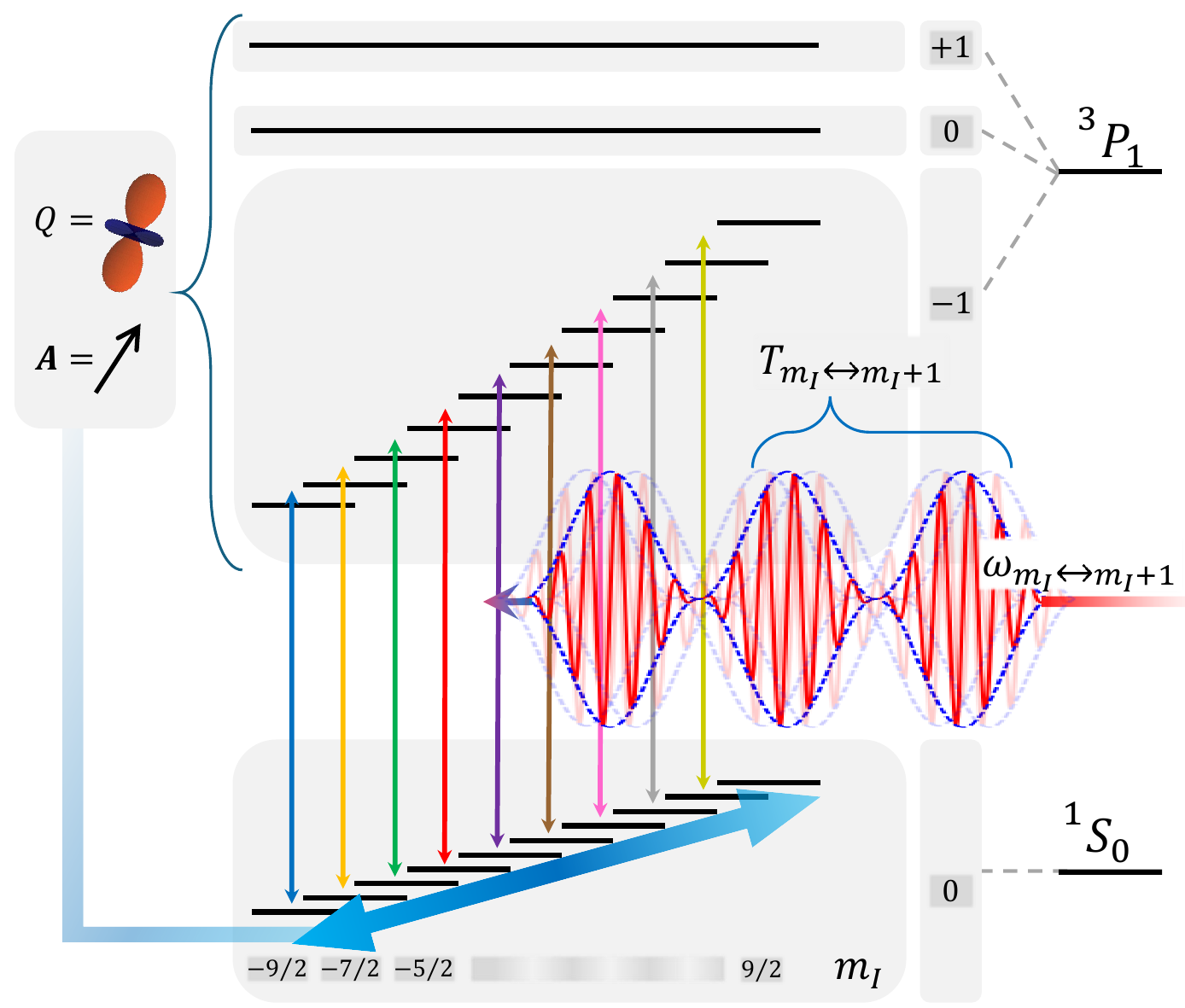}
    \caption{Schematic illustration of the optical nuclear electric resonance (ONER) protocol applied to the level structure of the (5s$^2$)~$^1S_0 \rightarrow$~(5s5p)~$^3P_1$ optical transition in $^{87}$Sr for qudit control. Amplitude-modulated laser fields with periods $T_{m_I\leftrightarrow m_I+1}$, frequencies $\omega_{m_I\leftrightarrow m_I+1}$ and electronic Rabi frequency $\Omega_\mathrm{E}/2\pi$ drive the system, resulting in an adiabatically modulated occupation of the electronically excited state. In the $^1S_0$ ground state, the magnetic nuclear spin quantum number $m_I$ remains well-defined. In the $^3P_1$ excited states, the non-zero hyperfine interaction ($Q,A \neq 0$) mixes nuclear spin states. Which states are mixed can be selected via the laser frequency $\omega_{m_I\leftrightarrow m_I+1}$, which is chosen to be in between the respective spin levels in the $m_J = -1$ manifold. The amplitude modulation is adjusted to the specific transition, enabling several hyperfine nuclear spin transitions in the $^1S_0$ ground state.}
    \label{fig:oner qudit schematic}
\end{figure}

We concentrate on the (5s$^2$)~$^1S_0 \rightarrow$~(5s5p)~$^3P_1$ transition in $^{87}$Sr for nuclear spin control in the ground state via the ONER method. We show that several one-level transitions of the ten-dimensional ground state manifold can be addressed by careful selection of the laser parameters. The principle of ONER for qudit control in $^{87}$Sr is illustrated in Figure~\ref{fig:oner qudit schematic}, where specific laser frequencies and amplitude modulations can drive different one-level transitions. In the case of single qubit manipulation, encoded in the $\Delta m_I=2$ transition between the $m_I=-9/2$ and $m_I =-5/2$ states,\cite{krondorfer2025opticalnuclearelectricresonance} ONER requires only moderate magnetic field strengths ($\gtrapprox$ 200 Gauss), just so that a Paschen-Back-like regime is entered for the states of interest in the hyperfine structure of the electronically excited state. However, for full qudit control within the same protocol, higher magnetic fields will be necessary to access multiple spin states with high fidelity, while simultaneously avoiding spin mixing. The relevant excited-state levels must be sufficiently spaced in energy, allowing the laser to be positioned between them while maintaining sufficient detuning from each transition.

\subsection{The $^1\text{S}_0\rightarrow\,^{3}\text{P}_1$ Transition in $^{87}\text{Sr}$}
For a theoretical investigation, we consider the Hamiltonian of a $^{87}$Sr atom in an external constant magnetic field $\bm{B}$ in $z$-direction, and an amplitude-modulated laser field with modulation period $T$ and polarization angle $\theta$ with respect to the $z$-axis within the rotating wave approximation (RWA). The amplitude modulation is set to $\Omega_\mathrm{E}(t) / 2\pi = \frac{\Omega_\mathrm{E} / 2\pi}{2} \left( 1 - \cos\left( \frac{2\pi t}{T} \right) \right)$. The total Hamiltonian is then given by
\begin{equation}\label{eq:full td Hamiltonian}
    H = H_\mathrm{E}' + H_\mathrm{Z} + H_\mathrm{HF} + H_\mathrm{AF}(t)\,,
\end{equation}
with the electronic Hamiltonian in the rotating frame $H_\mathrm{E}' = -\Delta\proj{^3P_1}$, with $\Delta = \omega_0 - \omega$, where $\omega_0$ is the central transition frequency of the (5s$^2$)~$^1S_0 \rightarrow$~(5s5p)~$^3P_1$ transition corresponding to 689 nm, and $\omega$ is the frequency of the laser field. Note that we set $\hbar = 1$ throughout the manuscript. The Zeeman Hamiltonian $H_\mathrm{Z}$ and the hyperfine Hamiltonian $H_\mathrm{HF}$ are given by
\begin{align}\label{eq:zeeman and hyperfine hamiltonian}
    \begin{split}
        H_\mathrm{Z} &= \left(g_J \mu_0 \hat{\bm{J}} - g_I \mu_\mathrm{N} \hat{\bm{I}} \right) \cdot \bm{B} \\
        H_\mathrm{HF} &= A \hat{\bm{I}}\cdot\hat{\bm{J}} + Q\frac{ \frac{3}{2}\hat{\bm{I}}\cdot\hat{\bm{J}} \left( 2\hat{\bm{I}}\cdot\hat{\bm{J}} + 1 \right) - \hat{\bm{I}}^2 \hat{\bm{J}}^2}{2IJ(2I-1)(2J-1)}\,,
    \end{split}
\end{align}
with the electronic total angular momentum $\hat{\bm{J}}$ and the nuclear spin $\hat{\bm{I}}$,  both multiplied by their corresponding magneton ($\mu_0$ and $\mu_\mathrm{N}$) and their g-factor, $g_J = 3/2$ and $g_I = -1.0928$, respectively.\cite{stone_table_2005} The hyperfine constants are $A = 2\pi \times -260\;\mathrm{MHz}$ and $Q = 2\pi \times -35\;\mathrm{MHz}$.\cite{zu_putlitz_bestimmung_1963} We choose a representation of the Hamiltonian in the basis $\ket{n,m_J,m_I}$, with $n$ denoting the electronic state, and $m_J$, $m_I$ denoting the magnetic quantum number of the total electronic angular momentum $\hat{\bm{J}}$ and nuclear spin~$\hat{\bm{I}}$.
In the dipole approximation, we then write the atom-field interaction Hamiltonian as
\begin{align}\label{eq:atom field hamiltonian}
\begin{split}
    H_\mathrm{AF}(t) = \frac{1}{2} \Omega_\mathrm{E}(t)\; ( 
    \cos(\theta) \ketbra{^1S_0}{^3P_1,0} &\otimes \mathbb{1}\,+ \\
    \frac{\sin(\theta)}{\sqrt{2}} \ketbra{^1S_0}{^3P_1,-1} &\otimes \mathbb{1}\,- \\
    \frac{\sin(\theta)}{\sqrt{2}} \ketbra{^1S_0}{^3P_1,+1} &\otimes \mathbb{1} + \text{h.c.} )\;,
\end{split}
\end{align}
with $\mathbb{1}$ denoting the unity matrix in the nuclear subspace. Additionally, we consider a decay from the $^3P_1$ state to the $^1S_0$ ground state with a decay rate of $\Gamma = 2\pi \times 7.48\;\mathrm{kHz}$.\cite{barnes_assembly_2022,heinz2020gamma}

\subsection{Simulation Details and Experimental Feasibility}
We simulate the full time-dependent Hamiltonian given in Equation~\eqref{eq:full td Hamiltonian} within the Python library QuTiP\cite{qutip1,qutip2} using the Lindblad master equation~\cite{breuer_opensystem_2007} with decay $\Gamma$ for the electronic levels. Via simulations of the full system, we
identify parameters that lead to a spin flip between the $m_I$ and $m_I + 1$ hyperfine ground states. For this purpose, we choose a laser frequency $\omega_{m_I\leftrightarrow m_I+1}$ exactly between the energies of the $\ket{^1S_0,0,m_I}\rightarrow\ket{^3P_1,-1,m_I}$ and $\ket{^1S_0,0,m_I+1}\rightarrow\ket{^3P_1,-1,m_I+1}$ transitions, i.e.
\begin{align}\label{eq: set laser frequency}
\begin{split}
    \omega&_{m_I\leftrightarrow m_I+1} = \frac{1}{2} ( ( E(^3P_1,-1,m_I) - E(^1S_0,0,m_I) ) \\ 
    &\quad+ (E(^3P_1,-1,m_I+1) - E(^1S_0,0,m_I+1)) )\;,
\end{split}
\end{align}
as also illustrated in Figure~\ref{fig:oner qudit schematic}.
This choice enhances individual control of one-level nuclear spin transitions and reduces coupling to other $\Delta m_I = \pm1$ transitions due to the detuning of the laser field from other transitions. Simultaneously, this choice of laser frequency ensures that $m_I$ and $m_I+1$ experience similar effective hyperfine interaction and show the same nuclear spin transition frequency. As illustrative example we choose a magnetic field of $B_0 = 3000\;\mathrm{G}$, a maximal electronic Rabi frequency $\Omega_{E,0} / 2\pi = 30\;\mathrm{MHz}$, a polarization angle $\theta_0 = 60\;\mathrm{deg}$, and the laser frequency $\omega_{m_I\leftrightarrow m_I+1}$, respectively the detuning $\Delta_{m_I \leftrightarrow m_I+1}$, as specified in Equation~\eqref{eq: set laser frequency}. The chosen electronic Rabi frequency is experimentally achievable; a detailed discussion can be found in Appendix~\ref{SI:dipole}. Extended scans over various parameter settings are also provided in the appendix. We note that rather large fields are necessary to obtain sufficient splitting of the relevant energy levels in the excited-state manifold. While such kilogauss-scale fields are above those typically used in neutral-atom experiments, high-precision $g$-factor measurements~\cite{Thekkeppatt2025}, Feshbach-resonance studies~\cite{Borowski2023bfield_stab}, and investigations of Rydberg spectra~\cite{Garton1980_rydberg} have been carried out in strontium at kilogauss fields, and ytterbium-based quantum computing experiments have also been performed in fields exceeding 500~G~\cite{norcia_midcircuit_2023}. This suggests that these field strengths are experimentally feasible and compatible with spin control on neutral atoms.

\section{Results}
\subsection{Nuclear Spin Transitions}
Scanning the amplitude modulation periods $T$, full flips of the nuclear spin typically appear for several equidistantly placed values of $T$.\cite{krondorfer2025opticalnuclearelectricresonance} Their actual position depends on the magnetic field $B_0$, the electronic Rabi frequency of the $^1S_0\rightarrow$$^3P_1$ transition $\Omega_\mathrm{E,0}/2\pi$, and the angle $\theta_0$. This periodicity is caused by multi-photon-like transitions, where the amplitude modulation period matches an effective energy difference between the target spin states,
\begin{equation}
    \Delta E_{m_I\leftrightarrow m_I+1}^\text{eff} = n\frac{2\pi}{T}\,,
\end{equation}
for some $n\in\mathbb{N}$. Due to the time-dependent light shifts generated by the amplitude-modulated laser field, an analytical computation of the effective energy difference is unfeasible. While in principle multiple transition peaks occur, we investigate the first peak for the respective parameters, as the highest Rabi frequency and the smoothest transitions occur at this peak.

\begin{figure*}[!ht]
    \centering
    \begin{subfigure}[b]{0.41\textwidth}
        \centering
        \includegraphics[width=\textwidth]{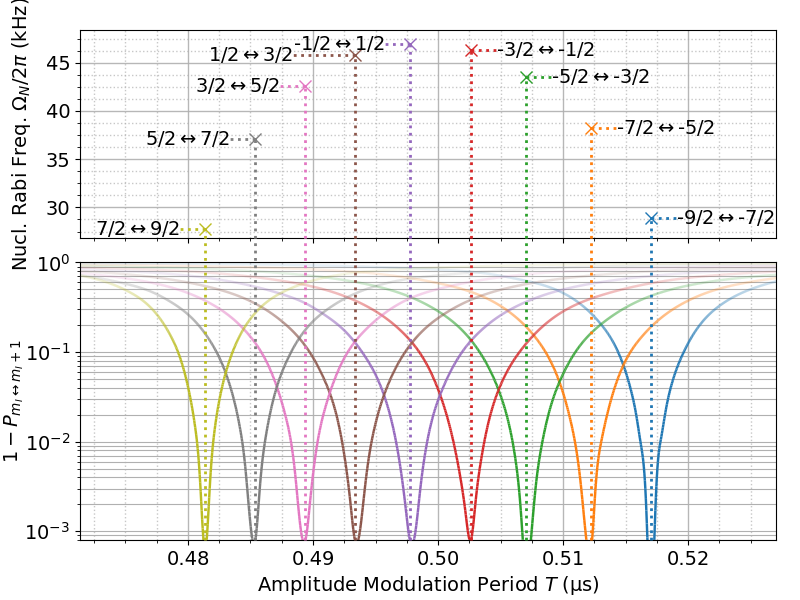}
        \put(-190,155){\small (a)}
        \phantomcaption\label{fig:ampl scan}
    \end{subfigure}%
    \begin{subfigure}[b]{0.59\textwidth}
        \centering
        \includegraphics[width=\textwidth]{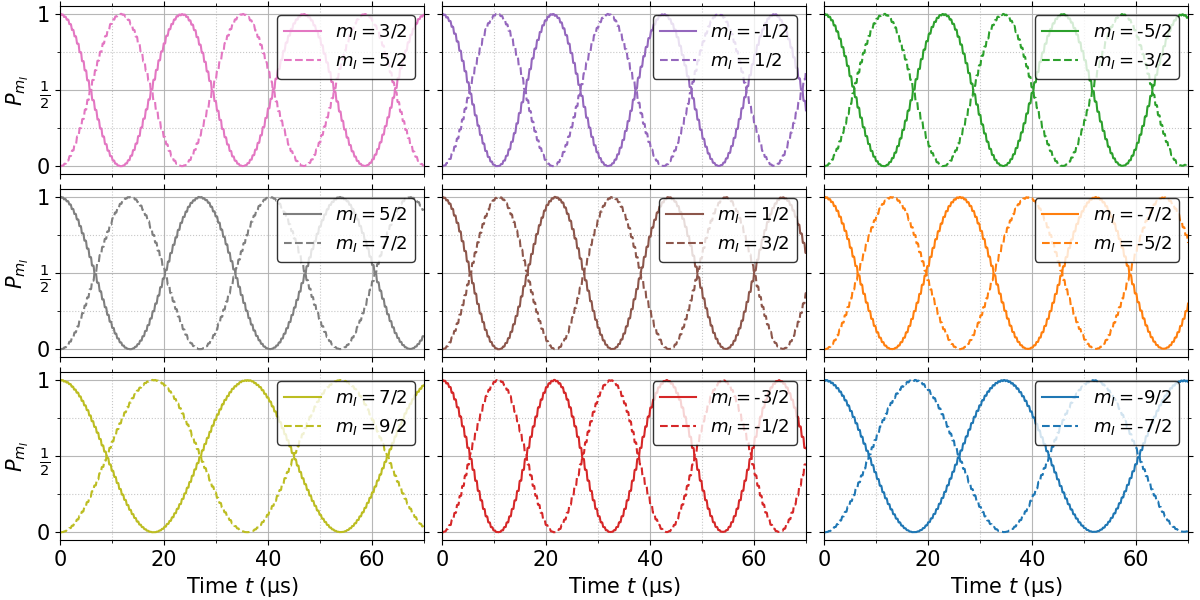}
        \put(-290,155){\small (b)}
        \phantomcaption\label{fig:rabi osci}
    \end{subfigure}
    \caption{(a) Amplitude modulation period scan and corresponding nuclear Rabi frequencies of $m_I \leftrightarrow m_I+1$ transitions. We show the $\pi$-pulse fidelity at the first flip $P_{m_I\leftrightarrow m_I+1}$ for all $\Delta m_I = \pm 1$ transitions at $B_0 = 3000\;\mathrm{G}$, $\Omega_{E,0} / 2\pi = 30\;\mathrm{MHz}$, $\theta_0 = 60\;\mathrm{deg}$ and $\Delta_{m_I \leftrightarrow m_I+1}$ as specified in Equation~\eqref{eq: set laser frequency}. Full transitions occur at different amplitude modulation periods $T_{m_I \leftrightarrow m_I+1}$ for different levels $m_I$. (b) Rabi oscillations of all $\Delta m_I = \pm 1$ at the optimal amplitude modulation period $T_{m_I \leftrightarrow m_I+1}$.} \label{fig:qudit transitions} 
\end{figure*}

In Figure~\ref{fig:ampl scan}, we present nuclear spin-flip probabilities (colored lines),
\begin{equation}
    P_{m_I \leftrightarrow m_I+1} = \max_{t\in\tau} \vert\braket{^1S_0,0,m_I+1 \vert \psi(t)}\vert^2\,,
\end{equation}
together with the resulting nuclear Rabi frequencies $\Omega_\mathrm{N}/2\pi$ (colored crosses) at the transition peaks, as a function of the modulation period~$T$. The probabilities are evaluated within a simulation interval $\tau = [0, 50\;\mathrm{\mu s}]$, where $\ket{\psi(t)}$ is the time-dependent wave function of the driven system, with $\ket{\psi(0)} = \ket{^1S_0,0,m_I}$.  Essentially, the flip probability $P_{m_I \leftrightarrow m_I+1}$ can be interpreted as the $\pi$-gate fidelity at the first flip for levels $m_I$ and $m_I+1$. Rabi oscillations at the maximum of the respective $m_I \leftrightarrow m_I+1$ transition are illustrated in Figure~\ref{fig:rabi osci}. For this parameter choice, all $\Delta m_I = \pm 1$ transitions can be independently addressed with high fidelity. Thus, ONER enables independent and high-fidelity control across the entire $^{87}$Sr nuclear spin manifold, a critical requirement for practical qudit quantum operations. The corresponding gate times, ranging from a few to tens of microseconds, ensure that these high-fidelity operations occur well within the typical coherence times of optical tweezer or lattice platforms, allowing the computational advantages of high-dimensional encoding to be realized at practical gate speeds.

For lower magnetic fields, however, only transitions of low-energy hyperfine ground states can be addressed with high fidelity. The reason for this limitation is spin mixing, which affects transitions to higher states if a sufficiently Paschen-Back-like regime is not reached, and the level splitting in the excited state manifold remains too small to allow a laser placement between the relevant states. Nevertheless, one-level transitions between the ground and first excited hyperfine ground state can still be addressed with high fidelity. To enable transitions at lower magnetic fields, $\Delta m_I = 2$ level transitions may be considered due to the larger level splitting. In this case, high fidelity control also at lower field strength becomes a possibility, but at the cost of a reduced (halved) number of levels available. A more detailed analysis of $\Delta m_I = 1$ transitions at different magnetic field strengths is provided in the Appendix~\ref{SI:other params}; a discussion of the lowest $\Delta m_I = 2$ transition is provided in Ref.~\citenum{krondorfer2025opticalnuclearelectricresonance}.

\subsection{Stability Analysis}
We now investigate the stability of the proposed ONER protocol for $m_I \rightarrow m_I+1$ qudit transitions in the electronic ground state for the reference parameters given above. The stability is checked with respect to the laser parameters $\theta$, $\Delta$ and $\Omega_\mathrm{E}$. Also, the robustness against variations in the magnetic field $B$ and the amplitude modulation period $T$ is tested. Concerning the role of heating and atom loss due to spontaneous scattering, we note that the simulation of the total system reveals that the occupation of the excited state is bounded by $P_{3P1} < 0.01$. The decay rate $\Gamma$ for the $^3P_1$ state of $^{87}$Sr leads to 
\begin{equation}
    N_{\mathrm{sc}} = \frac{P_{3P1} \; \Gamma}{\Omega_\mathrm{N}} < 0.004
\end{equation}
as the total number of spontaneously scattered photons per Rabi cycle, assuming a nuclear Rabi frequency of $\Omega_\mathrm{N}/2\pi \gtrapprox 20\;\mathrm{kHz}$. Therefore, we can neglect heating and atom loss in this case.

To evaluate the stability with respect to the laser parameters and the magnetic field, we perform simulations with perturbed parameters and investigate the transition probability of the respective target states. We simulate the time evolution and calculate the $\pi$-pulse fidelity of the $\ket{^1S_0,0,m_I}\leftrightarrow\ket{^1S_0,0,m_I+1}$ transition under perturbed parameters. The simulation results are summarized in Table~\ref{tab:tolerance}, where we show the minimum fidelity over all one-level transitions,
\begin{equation}
P_{\min} = \min_{m_I}P_{m_I\leftrightarrow m_I+1}\,.
\end{equation}
Thus, if the deviation of each parameter is below the given threshold value for the target fidelity of 99\% or 99.9\%, all $\Delta m_I = \pm 1$ transitions can be addressed with the respective fidelity.

\begin{table}[!htb]
\caption{Tolerances for qudit control with ONER at $B_0 = 3000$ G, $\Omega_{E,0} / 2\pi = 30$ MHz, $\theta_0 = 60$ deg, and the corresponding $\Delta_{m_I \leftrightarrow m_I+1}$ and $T_{m_I \leftrightarrow m_I+1}$. We show tolerances for the minimal flip probability $P_{\min}$, both for 99\% and 99.9\% target fidelities. Deviations smaller than the listed tolerances yield spin-flip control for all one-level transitions with the respective fidelity.}
\label{tab:tolerance}
\centering
\begin{tabular}{@{}c|ccccc@{}}
\toprule
& $\abs{\delta T}$ & $\abs{\delta B}$ & $\abs{\delta\Omega_E/2\pi}$ & $\abs{\delta\theta}$ & $\abs{\delta\Delta}$ \\
&  (ns) & (mG) & (kHz) & (deg) & (kHz) \\
\midrule
99.9\% & 0.1 & 300 & 8 & 0.03 & 500 \\
99\%   & 0.6 & 3000 & 80 & 0.2 & 5000 \\
\bottomrule
\end{tabular}
\end{table}

Noise in the magnetic field up to 3~G ($\sim$0.1\% for $B_0 = 3000\;\mathrm{G}$) is tolerable for a $\pi$-pulse fidelity exceeding 99\%. These tolerances are readily achieved with commercial power supplies and can be exceeded with modern stabilization techniques.\cite{Borowski2023bfield_stab} Similar fidelities can be achieved with detunings from the optimal laser frequency of up to 5~MHz, a polarization angle perturbation of up to 0.2 degrees, and perturbations in the amplitude modulation period of up to 0.6~ns ($\sim$0.1\% for $T_{m_I,0} \approx 0.5\;\mathrm{\mu s}$). Noise in the electronic Rabi frequency $\Omega_\mathrm{E}$ is tolerable up to 0.25\%, equivalent to about 80~kHz. This translates into tolerable intensity noise of 0.5\%, by using the relation $\Omega_\mathrm{E} \sim \sqrt{\mathcal{I}}$ and thus $\delta\Omega_\mathrm{E} / \Omega_\mathrm{E} = \frac{1}{2}\delta \mathcal{I} / \mathcal{I}$. Commercial stabilization systems readily achieve an intensity stability of 0.05\%, which corresponds to 0.025\% in electronic Rabi frequency. Details on the relation between electronic Rabi frequency and laser intensity are provided in the Appendix~\ref{SI:dipole}. For small variations and fine-tuned control parameters, also fidelities exceeding 99.9\% are possible. Thus, we observe stable results with respect to magnetic field perturbations, detuning, polarization angle, amplitude modulation period, and intensity, for experimentally reasonable parameter values and variations.

The stability analysis performed above shows the dependence of the tuning parameters and the stability with respect to quasi-static variations. The analysis remains valid as long as the timescale of the variations is large compared to the timescale of the gate operation. If this is not the case, the time-dependence of the variations has to be described with a suitable time-dependent noise model, fitted to the experimentally observable variations of the individual parameters.

\section{Conclusion and Outlook}
In summary, we demonstrated that optical nuclear electric resonance (ONER) is a powerful method for rapid and robust single-qudit control in trapped neutral atoms. By selectively addressing multiple one-level nuclear spin transitions with tailored amplitude-modulated laser fields, we achieved fast qudit operations with high fidelity even in the presence of typical noise sources. This technique provides a compelling alternative to methods based on radio frequencies~\cite{omanakuttan_qudit_2023,Deutsch2021qudit} and standard two-photon Raman transitions~\cite{barnes_assembly_2022,ahmed_coherent_2025}, by achieving higher nuclear Rabi frequencies and providing improved spatial localization, compared to magnetic control. While we focused our analysis on $^{87}$Sr, which offers a total of $d=10$ spin states, the method is equally applicable to other alkaline earth and alkaline earth-like atoms. Note that a ten-dimensional Hilbert space translates into $\log_2(10)\!\approx\!3.3$ effective qubits. This higher information density reduces circuit width and depth by factors of about three to ten, depending on algorithmic scaling~\cite{luo_universal_2014, ringbauer_universal_2022}. With ONER-based single-qudit Rabi rates of tens of kHz, these reductions suggest faster circuit execution on experimentally realistic timescales. Full access to SU($N$) with $N\!\le\!10$ enables compact implementations of error-correcting codes that can tolerate up to $(d\!-\!2)/2\!=\!4$ level errors within a single atom~\cite{Nikolaeva2024}, and native realization of higher-spin Hamiltonians~\cite{Zhang2014}. Paired with non-destructive, state-resolved readout and two-qudit interactions \--- both compatible with ONER \--- this provides a concrete and experimentally viable route towards utilizing the computational advantages of high-dimensional qudits in neutral-atom systems.

Even though large magnetic fields are required, ONER should remain compatible with two-qudit gate mechanisms. At the time of writing, no direct experimental studies on Rydberg interactions in strontium at kilogauss fields exist. Still, prior work shows that Rydberg spectra and interactions remain tractable even under strong fields~\cite{Pohl2009,Garton1980_rydberg,Neukammer1984,Robicheaux2018}, and recent Yb experiments have demonstrated blockade at several hundred Gauss~\cite{norcia_midcircuit_2023}, suggesting that the extension to higher fields is realistic. Alternative schemes, such as controlled collisions~\cite{Mandel2003}, superexchange~\cite{Trotzky2008}, or SU$(N)$-symmetric interactions in Sr~\cite{Zhang2014}, might also provide viable routes for two-qudit gates.

Among the possible pathways for realizing two-qudit interactions, Rydberg blockade appears particularly promising in combination with ONER. The excitation scheme commonly used for neutral-atom qubits, via the $^3P_1$ or $^3P_2$ intermediate states into high-lying $n\,^3S_1$ Rydberg levels, extends to qudits~\cite{omanakuttan_qudit_2023} and should remain applicable even in the presence of large magnetic fields. Studies of strong-field Rydberg spectra in alkaline-earth systems suggest that such operation is feasible~\cite{Garton1980_rydberg, Pohl2009, Neukammer1984}, and the expected Rydberg transition rates of several megahertz would readily support fast, high-fidelity conditional-phase gates. In the simplest implementation, only a single, selected control level (e.g., $m_I=-9/2$) may be optically coupled to a Rydberg state, while all other nuclear-spin levels remain off-resonant. This state-selective Rydberg excitation produces a conditional phase (CZ) operation between two qudits, and additional single-qudit rotations can compensate residual phases in the remaining sublevels. Together with universal single-qudit control provided by ONER, such an entangling operation is sufficient for universal qudit computation~\cite{Brylinski2002}.

Alternative schemes employing multiple Rydberg transitions to address different control states may also be feasible, provided that the relevant transitions can be spectrally separated. In either case, the achievable interaction strengths are comparable to those demonstrated in qubit-based Rydberg systems, indicating that ONER-assisted two-qudit gates could, in principle, operate on similar timescales. A detailed investigation of Rydberg blockade and level mixing in the high-field regime will be an important next step toward a quantitative assessment of this approach.

High-fidelity readout is likewise feasible: large Zeeman splittings in the excited state manifold enable state-selective fluorescence on the $^1S_0 \rightarrow ^3P_1$ transition, as demonstrated in Yb arrays~\cite{norcia_midcircuit_2023}, while shelving to $^3P_0$ or $^3P_2$ combined with blow-away detection~\cite{saffman_quantum_2016} provides a complementary destructive read-out approach. Moreover, large magnetic fields are not only compatible with typical cooling schemes but might even enable spin-coherent cooling.\cite{Reichenbach2007_cooling_strong, Shi2023_cooling_weak}

This establishes ONER as a promising approach for scalable high-dimensional quantum computing, reducing circuit complexity and improving algorithmic efficiency. Given its robustness and experimental feasibility, the protocol opens new opportunities for qudit-based quantum error correction, high-dimensional entanglement, and quantum simulation. Future work will focus on lowering the required magnetic fields, extending ONER-based schemes to multi-qudit entanglement operations, and exploring their integration into large-scale neutral-atom quantum processors. Overall, ONER paves the way toward fast, coherent, and scalable high-dimensional quantum information processing.

\bigskip
\begin{appendix}
\section{Overview}
In this appendix, we provide an extended scan of the amplitude modulation $T$ to illustrate the multiple equidistant transition peaks in Appendix~\ref{SI:ampl scan}. We provide scans of the amplitude modulation period for different system parameters in Appendix~\ref{SI:other params} and discuss resulting Rabi oscillations in more detail. Finally, details on the overall computational implementation are presented in Appendix~\ref{SI:comp}, while details on the necessary laser intensity can be found in Appendix~\ref{SI:dipole}

\section{Extended $T$ Scan -- Multiple Transitions}\label{SI:ampl scan}
\begin{figure*}[!htb]
    \centering
    \includegraphics[width=\textwidth]{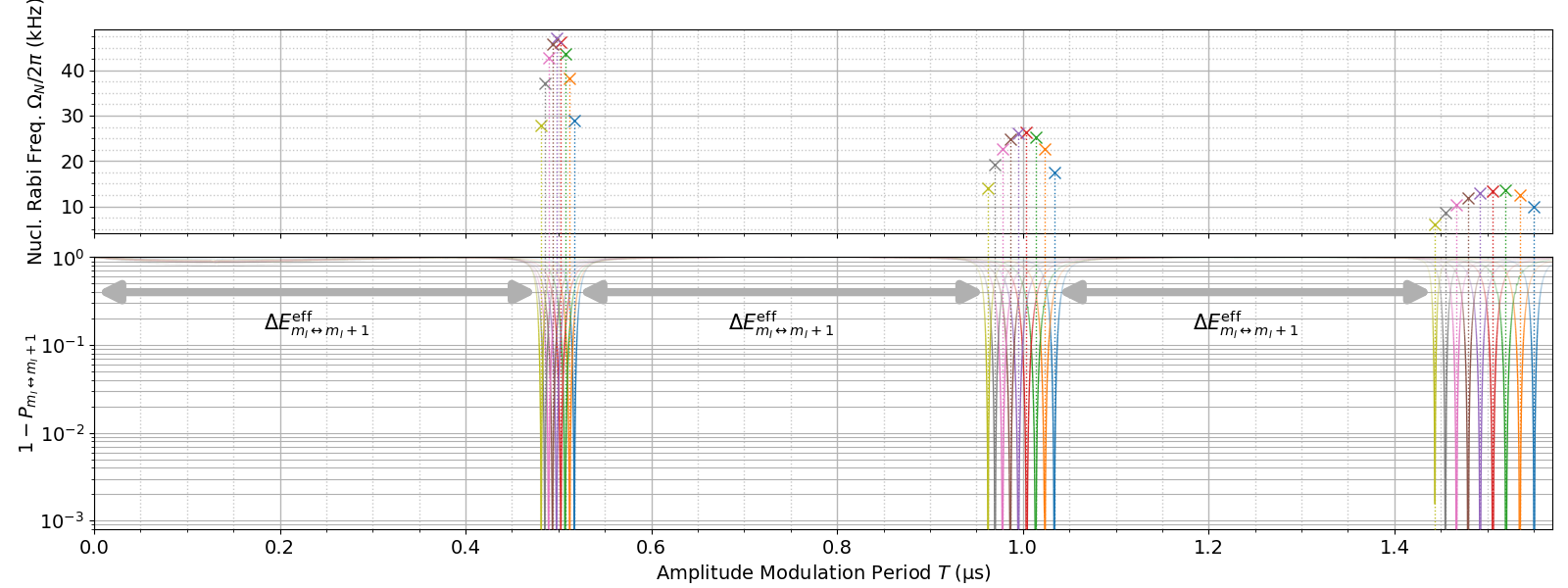}
    \caption{Amplitude modulation period scan and corresponding nuclear Rabi frequencies. We show the $\pi$-pulse fidelity at the first flip $P_{m_I\leftrightarrow m_I+1}$ for all $\Delta m_I = \pm 1$ transitions at parameters $B_0 = 3000\;\mathrm{G}$, $\Omega_{E,0} / 2\pi = 30\;\mathrm{MHz}$, $\theta_0 = 60\;\mathrm{deg}$ and $\omega_{m_I \leftrightarrow m_I+1}$ as specified in the main text. Full transitions occur at different amplitude modulation periods $T_{m_I \leftrightarrow m_I+1}$ for different levels $m_I$. Moreover, multiple equidistantly spaced transitions occur for each level.}\label{fig:ampl scan extended} 
\end{figure*}

In this section, we show that full transitions occur for equidistantly spaced amplitude modulation periods $T$, for all one-level transitions. As explained in the main text, this is due to multi-photon-like transitions that obey the equation
\begin{equation}
    \Delta E_{m_I\leftrightarrow m_I+1}^\text{eff} = n\frac{2\pi}{T}\,,
\end{equation}
with $n\in\mathbb{N}$. The effective energy difference $\Delta E_{m_I\leftrightarrow m_I+1}^\text{eff}$ of the ground state hyperfine levels is state dependent, so the periodicity is different for different transitions $m_I\leftrightarrow m_I+1$. Figure~\ref{fig:ampl scan extended} shows the multiple transition peaks for the parameters $B_0 = 3000\;\mathrm{G}$, a maximal electronic Rabi frequency $\Omega_{E,0} / 2\pi = 30\;\mathrm{MHz}$, a polarization angle $\theta_0 = 60\;\mathrm{deg}$, and the laser frequency $\omega_{m_I\leftrightarrow m_I+1}$. We can see that the nuclear Rabi frequency approximately drops in a straight line for higher-order transition peaks.

\section{$T$ Scan for Other Parameters}\label{SI:other params}
\begin{figure*}[!htb]
    \centering
    \begin{subfigure}[b]{0.41\textwidth}
        \centering
        \includegraphics[width=\textwidth]{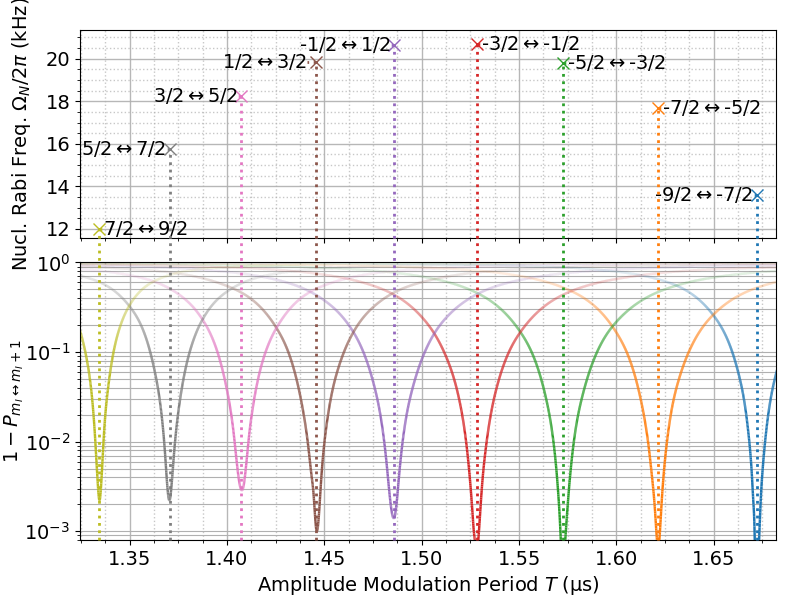}
        \put(-190,155){\small (a)}
        \phantomcaption\label{fig:ampl scan 1000}
    \end{subfigure}%
    \begin{subfigure}[b]{0.59\textwidth}
        \centering
        \includegraphics[width=\textwidth]{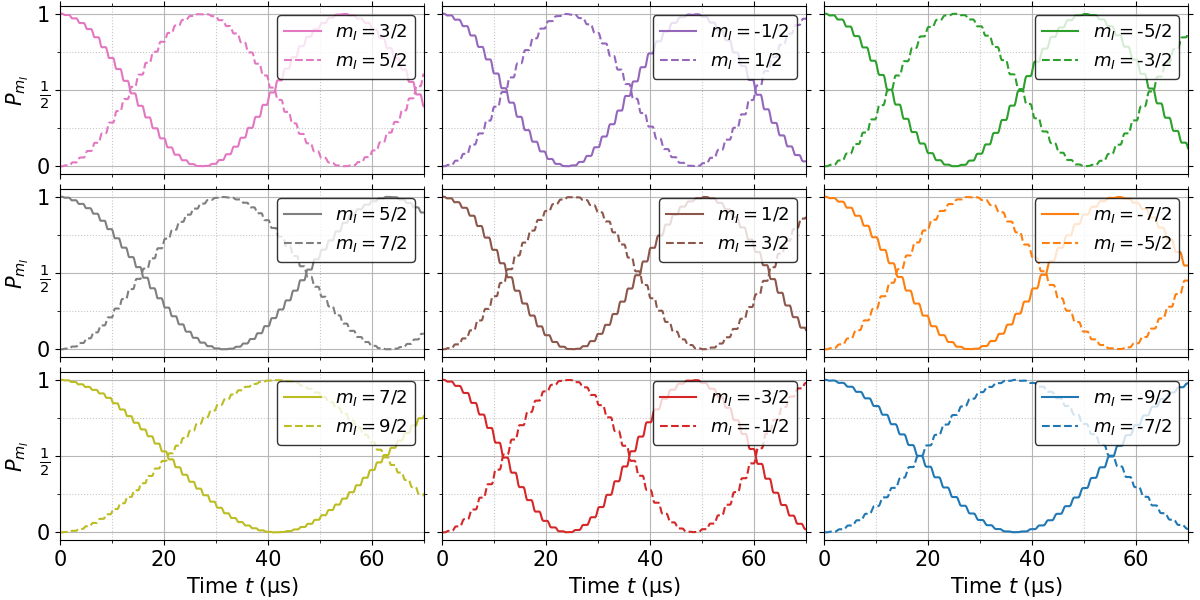}
        \put(-290,155){\small (b)}
        \phantomcaption\label{fig:rabi osci 1000}
    \end{subfigure}
    \caption{(a) Amplitude modulation period scan and corresponding nuclear Rabi frequencies. We show the $\pi$-pulse fidelity at the first flip $P_{m_I\leftrightarrow m_I+1}$ for all $\Delta m_I = \pm 1$ transitions at parameters $B_0 = 1000\;\mathrm{G}$, $\Omega_{E,0} / 2\pi = 15\;\mathrm{MHz}$, $\theta_0 = 75\;\mathrm{deg}$ and $\omega_{m_I \leftrightarrow m_I+1}$ as specified in the main text. Full transitions occur at different amplitude modulation periods $T_{m_I \leftrightarrow m_I+1}$ for different levels $m_I$. (b) Rabi oscillations of all $\Delta m_I = \pm 1$ at the optimal amplitude modulation period $T_{m_I \leftrightarrow m_I+1}$.} \label{fig:qudit transitions 1000} 
\end{figure*}

\begin{figure*}[!htb]
    \centering
    \begin{subfigure}[b]{0.41\textwidth}
        \centering
        \includegraphics[width=\textwidth]{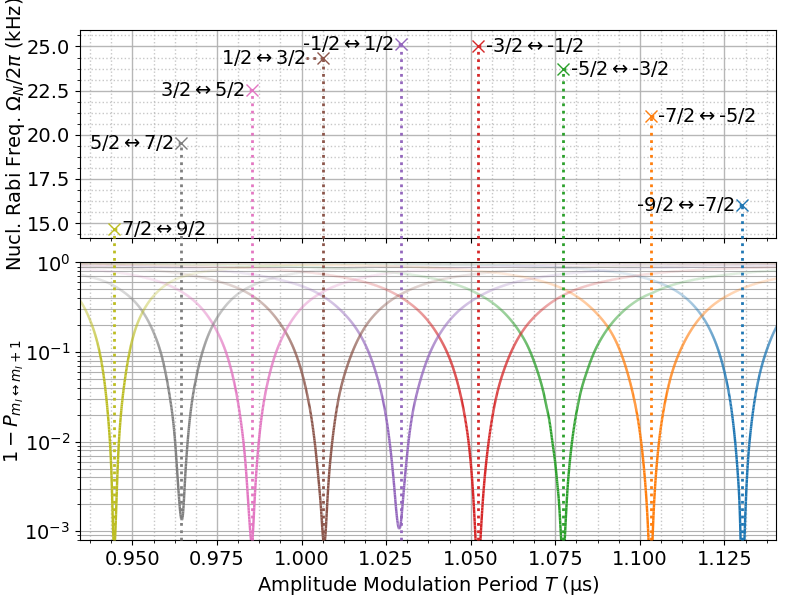}
        \put(-190,155){\small (a)}
        \phantomcaption\label{fig:ampl scan 1500}
    \end{subfigure}%
    \begin{subfigure}[b]{0.59\textwidth}
        \centering
        \includegraphics[width=\textwidth]{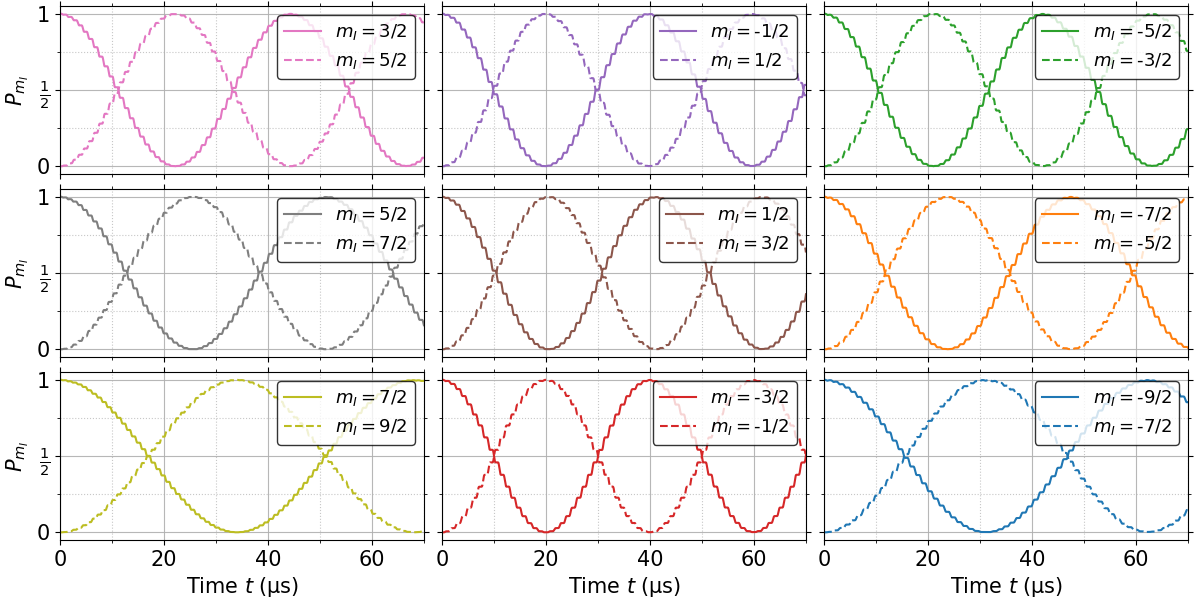}
        \put(-290,155){\small (b)}
        \phantomcaption\label{fig:rabi osci 1500}
    \end{subfigure}
    \caption{(a) Amplitude modulation period scan and corresponding nuclear Rabi frequencies. We show the $\pi$-pulse fidelity at the first flip $P_{m_I\leftrightarrow m_I+1}$ for all $\Delta m_I = \pm 1$ transitions at parameters $B_0 = 1500\;\mathrm{G}$, $\Omega_{E,0} / 2\pi = 20\;\mathrm{MHz}$, $\theta_0 = 75\;\mathrm{deg}$ and $\omega_{m_I \leftrightarrow m_I+1}$ as specified in the main text. Full transitions occur at different amplitude modulation periods $T_{m_I \leftrightarrow m_I+1}$ for different levels $m_I$. (b) Rabi oscillations of all $\Delta m_I = \pm 1$ at the optimal amplitude modulation period $T_{m_I \leftrightarrow m_I+1}$.} \label{fig:qudit transitions 1500} 
\end{figure*}

The main text is dedicated to transitions for $B_0 = 3000\;\mathrm{G}$, a maximal electronic Rabi frequency $\Omega_{E,0} / 2\pi = 30\;\mathrm{MHz}$, a polarization angle $\theta_0 = 60\;\mathrm{deg}$, and the laser frequency $\omega_{m_I\leftrightarrow m_I+1}$, respectively the detuning $\Delta_{m_I \leftrightarrow m_I+1}$, since all one-level transition can be addressed with a fidelity exceeding 99.9\%. However, the ONER method is also applicable at lower magnetic fields. In these cases, not all one-level transitions may be addressed with a fidelity this high -- but a subset remains usable. Therefore, it makes sense to look at scans over various system parameters for these scenarios as well. Again, we focus on one-level transitions, in contrast to Ref.~\citenum{krondorfer2025opticalnuclearelectricresonance}, where two-level transitions have been considered. We investigate the system parameters $B_0 = 1000\;\mathrm{G}$, $\Omega_{E,0} / 2\pi = 15\;\mathrm{MHz}$, $\theta_0 = 75\;\mathrm{deg}$, and the corresponding $\Delta_{m_I \leftrightarrow m_I+1}$ in Figure~\ref{fig:qudit transitions 1000}. In Figure~\ref{fig:qudit transitions 1500} we look at the transitions for system parameters $B_0 = 1500\;\mathrm{G}$, $\Omega_{E,0} / 2\pi = 20\;\mathrm{MHz}$, $\theta_0 = 75\;\mathrm{deg}$, and the corresponding $\Delta_{m_I \leftrightarrow m_I+1}$.

In both cases, we obtain fidelities exceeding 99.9\% for the lower four levels. For higher levels, however, the fidelity is quenched, but still exceeding 99\%, as illustrated in Figures~\ref{fig:ampl scan 1000}~and~\ref{fig:ampl scan 1500}. Thresholds, determined by a scan over small variations of the system parameters, are provided in Table~\ref{tab:tolerance 1000} and Table~\ref{tab:tolerance 1500} for each individual transition. In both cases, high fidelity transitions can still be reached for reasonable parameters and tolerance thresholds. Note that the given thresholds are estimated from below, based on the simulated transition probability with perturbed parameters. Larger perturbations might therefore still yield accurate results, especially since different parameters can be co-dependent. This is analyzed in more detail in Ref.~\citenum{krondorfer2025opticalnuclearelectricresonance}.

\begin{table*}[!htb]
\caption{Tolerances for qudit control with ONER at $B_0 = 1000\;\mathrm{G}$, $\Omega_{E,0} / 2\pi = 15\;\mathrm{MHz}$, $\theta_0 = 75\;\mathrm{deg}$, and the corresponding $\Delta_{m_I \leftrightarrow m_I+1}$ and $T_{m_I \leftrightarrow m_I+1}$. We show tolerances individually for each $\Delta m_I = \pm 1$ transition, both for 99\% and 99.9\% target fidelities. Deviations smaller than the listed tolerances yield spin-flip control for the corresponding transition with the respective fidelity.}
\label{tab:tolerance 1000}
\centering
\rowcolors{2}{gray!10}{white}
\begin{tabular}{c|c|c|c|c|c|c|c|c|c|c}
\toprule
 & \multicolumn{2}{c|}{$\abs{\delta T}$ (ns)} 
 & \multicolumn{2}{c|}{$\abs{\delta B}$ (mG)}
 & \multicolumn{2}{c|}{$\abs{\delta\Omega_E/2\pi}$ (kHz)}
 & \multicolumn{2}{c|}{$\abs{\delta\theta}$ (deg)}
 & \multicolumn{2}{c}{$\abs{\delta\Delta}$ (kHz)} \\
\midrule
\textbf{Transition} & 99.9\% & 99\% & 99.9\% & 99\% & 99.9\% & 99\% & 99.9\% & 99\% & 99.9\% & 99\% \\
\midrule
$-9/2 \leftrightarrow -7/2$ & 0.8 & 3.5 & 380 & $>$1000 & 10 & 35 & 0.15 & 0.7 & $>$5000 & $>$5000 \\
$-7/2 \leftrightarrow -5/2$ & 0.8 & 3.5 & 380 & $>$1000 & 10 & 60 & 0.15 & 0.9 & $>$5000 & $>$5000 \\
$-5/2 \leftrightarrow -3/2$ & 0.8 & 4.0 & 400 & $>$1000 & 20 & 60 & 0.15 & 0.9 & $>$5000 & $>$5000 \\
$-3/2 \leftrightarrow -1/2$ & 0.8 & 4.0 & 400 & $>$1000 & 15 & 80 & 0.15 & 1.0 & $>$5000 & $>$5000 \\
$-1/2 \leftrightarrow 1/2$ & - & 3.8 & - & $>$1000 & - & 80 & - & 1.0 & - & $>$5000 \\
$1/2 \leftrightarrow 3/2$ & - & 3.8 & - & $>$1000 & - & 80 & - & 1.0 & - & $>$5000 \\
$3/2 \leftrightarrow 5/2$ & - & 3.0 & - & $>$1000 & - & 80 & - & 1.0 & - & $>$5000 \\
$5/2 \leftrightarrow 7/2$ & - & 2.5 & - & $>$1000 & - & 80 & - & 1.0 & - & $>$5000 \\
$7/2 \leftrightarrow 9/2$ & - & 2.0 & - & $>$1000 & - & 80 & - & 1.0 & - & $>$5000 \\
\bottomrule
\end{tabular}
\end{table*}

\begin{table*}[!htb]
\caption{Tolerances for qudit control with ONER at $B_0 = 1500\;\mathrm{G}$, $\Omega_{E,0} / 2\pi = 20\;\mathrm{MHz}$, $\theta_0 = 75\;\mathrm{deg}$, and the corresponding $\Delta_{m_I \leftrightarrow m_I+1}$ and $T_{m_I \leftrightarrow m_I+1}$. We show tolerances individually for each $\Delta m_I = \pm 1$ transition, both for 99\% and 99.9\% target fidelities. Deviations smaller than the listed tolerances yield spin-flip control for the corresponding transition with the respective fidelity.}
\label{tab:tolerance 1500}
\centering
\rowcolors{2}{gray!10}{white}
\begin{tabular}{c|c|c|c|c|c|c|c|c|c|c}
\toprule
 & \multicolumn{2}{c|}{$\abs{\delta T}$ (ns)} 
 & \multicolumn{2}{c|}{$\abs{\delta B}$ (mG)}
 & \multicolumn{2}{c|}{$\abs{\delta\Omega_E/2\pi}$ (kHz)}
 & \multicolumn{2}{c|}{$\abs{\delta\theta}$ (deg)}
 & \multicolumn{2}{c}{$\abs{\delta\Delta}$ (kHz)} \\
\midrule
\textbf{Transition} & 99.9\% & 99\% & 99.9\% & 99\% & 99.9\% & 99\% & 99.9\% & 99\% & 99.9\% & 99\% \\
\midrule
$-9/2 \leftrightarrow -7/2$ & 0.5 & 1.8 & 350 & $>$1000 & 8 & 40 & 0.10 & 0.45 & $>$5000  & $>$5000  \\
$-7/2 \leftrightarrow -5/2$ & 0.5 & 2.0 & 450 & $>$1000 & 10 & 50 & 0.10 & 0.60 & $>$5000  & $>$5000  \\
$-5/2 \leftrightarrow -3/2$ & 0.5 & 2.0 & 450 & $>$1000 & 15 & 70 & 0.10 & 0.70 & $>$5000  & $>$5000  \\
$-3/2 \leftrightarrow -1/2$ & 0.5 & 2.5 & 450 & $>$1000 & 20 & 70 & 0.15 & 0.80 & $>$5000  & $>$5000  \\
$-1/2 \leftrightarrow 1/2$ & - & 2.0 & - & $>$1000 & - & 70 & - & 0.80 & - & $>$5000  \\
$1/2 \leftrightarrow 3/2$ & - & 2.0 & - & $>$1000 & - & 70 & - & 0.80 & - & $>$5000  \\
$3/2 \leftrightarrow 5/2$ & - & 1.8 & - & $>$1000 & - & 70 & - & 0.80 & - & $>$5000  \\
$5/2 \leftrightarrow 7/2$ & - & 1.8 & - & $>$1000 & - & 70 & - & 0.80 & - & $>$5000  \\
$7/2 \leftrightarrow 9/2$ & - & 1.4 & - & $>$1000 & - & 70 & - & 0.60 & - & $>$5000  \\
\bottomrule
\end{tabular}
\end{table*}

For both parameter sets, nuclear Rabi frequencies between 10 and 25~kHz are achieved. The corresponding Rabi oscillations are illustrated in Figures~\ref{fig:rabi osci 1000}~and~\ref{fig:rabi osci 1500}. The Rabi oscillations display a step-like behavior, which stems from the amplitude modulation: Each step corresponds to one cycle in the periodic modulation of the amplitude. In order to obtain smooth Rabi oscillations, the amplitude modulation frequency should be as high as possible.

\section{Computational Details}\label{SI:comp}
As a first step, the total time-dependent Hamiltonian is set up according to Equation~(1) of the main text. 
As basis states we choose $\ket{n,m_J,m_I}$, with $n\in\{^1S_0,^3P_1\}$, $m_J = 0$ for $^1S_0$ state and $m_J\in\{\pm 1, 0\}$ for $^3P_1$ state and $m_I\in\{-9/2,-7/2,...,9/2\}$ as discussed in the main text. Initial states are chosen as
\begin{equation}
    \ket{\psi_0}\in\left\{ \ket{^1S_0,0,m_I}\,\vert\, m_I = -9/2, \dots, 9/2\right\}\,.
\end{equation}
Since the excited $^3P_1$ state decays with rate $\Gamma = 2\pi*7.48\;\mathrm{kHz}$ into the $^1S_0$ ground state,\cite{barnes_assembly_2022,heinz2020gamma} an open quantum system has to be considered with density matrix $\rho$. The time-evolution of the system can then be described by a Born-Markov Master equation~\cite{breuer_opensystem_2007} of the form
\begin{equation} \label{eq:born markov}
    \rmi\hbar\partial_t \rho = \com{H}{\rho} + \rmi\hbar\sum_\alpha k_\alpha \mathcal{L}[c_\alpha]\rho_S,
\end{equation}
with decay strengths $k_\alpha = \Gamma$ and the Lindblad superoperator $\mathcal{L}$ defined by
\begin{equation}
    \mathcal{L}[c]\rho_S = c\rho_S c^\dagger - \frac{1}{2}\left( c^\dagger c \rho_S + \rho_S c^\dagger c \right)
\end{equation}
for a collapse operator $c$. In our case, the collapse operators are given by
\begin{align}
    \begin{split}
        c_0 &= \ket{^1S_0,0}\bra{^3P_1,0} \otimes \mathbb{1}\;, \\
        c_+ &= \ket{^1S_0,0}\bra{^3P_1,+1} \otimes \mathbb{1}\;, \\
        c_- &= \ket{^1S_0,0}\bra{^3P_1,-1} \otimes \mathbb{1}\;,
    \end{split}
\end{align}
and model the spontaneous decay from the excited state to the ground state under the emission of a linearly, right- or left-circularly polarized photon. We simulate the time evolution given by Equation~\eqref{eq:born markov} in a time interval $\tau$ with the Python library QuTiP~\cite{qutip1,qutip2}, using the \textit{mesolve} function. From the simulation results, we extract the occupation of the respective states.

\section{Conversion of Rabi frequency and laser intensity}\label{SI:dipole}  
For the ONER protocol, the electronic Rabi frequency $\Omega_\mathrm{E}/2\pi$ of the $^{1}S_{0}\rightarrow{}^{3}P_{1}$ transition in $^{87}$Sr provides a convenient measure of the required laser intensity. The relation between Rabi frequency and intensity is determined by the dipole matrix element of the transition,  
\[
\mathcal{D} = \left|\langle {}^{1}S_{0}|\hat{\bm{d}}|{}^{3}P_{1}\rangle\right| = \frac{0.151}{\sqrt{3}}~\text{a.u.},
\]  
together with a Wigner factor of $1/\sqrt{3}$~\cite{cooper2018dipole}.  
The connection between laser intensity $\mathcal{I}$ and Rabi frequency follows from  
\[
\hbar \Omega_\mathrm{E} = -\langle {}^{1}S_{0}|\hat{\bm{d}}|{}^{3}P_{1}\rangle \cdot \bm{E}_0,
\]  
with $\mathcal{I} = \tfrac{1}{2}\varepsilon_{0}cE_{0}^{2}$. This gives  
\begin{equation}
    \left|\hbar \Omega_\mathrm{E}\right| = \mathcal{D}\,\sqrt{\tfrac{2\mathcal{I}}{\varepsilon_{0}c}}.
\end{equation}  

From this relation, electronic Rabi frequencies of $\Omega_\mathrm{E}/2\pi = 20,\,40,\,60~\text{MHz}$ correspond to intensities of $\mathcal{I} = 1,\,4,\,10~\text{W/cm}^{2}$, respectively. For a Gaussian beam of waist $w_{0}$, the peak intensity is given by  
\(\mathcal{I} = \frac{2P}{\pi w_{0}^{2}},\)
where $P$ is the optical power. Thus, the above intensities translate to a laser power of several microwatts to a few milliwatts for beam diameters between $20$ and $100~\mu\text{m}$.

These values show that MHz-scale Rabi frequencies are readily achievable with only tens of microwatts to a few milliwatts of optical power, which is well within the reach of current strontium tweezer and lattice platforms. A~similar discussion is provided in Ref.~\citenum{krondorfer2025opticalnuclearelectricresonance}.

\end{appendix}

\bigskip
\begin{acknowledgements}
We acknowledge funding by the Austrian Science Fund (FWF) [10.55776/P36903]. We further thank the IT Services (ZID) of the Graz University of Technology for providing high-performance computing resources and technical support.
\end{acknowledgements}

\input{output.bbl}

\end{document}

%% file: output.bbl
%